\documentclass{article}
\begin{document}

\begin{center}{\large \bf About nondegenerative four photon mixing gain 
spectroscopy of  the $SF_6$ molecules.}
\end{center}

\vspace{3 pt}
\centerline{\sl V.A.Kuz'menko}
\vspace{5 pt}
\centerline{\small \it Troitsk Institute of Innovation and Fusion Research,}
\centerline{\small \it Troitsk, Moscow region, 142092, Russian Federation.}
\vspace{5 pt}
\begin{abstract}

Infrared-infrared double resonance measurements of the $\nu_3$ absorption 
band of $SF_6$ using a $CO_2$ TEA laser radiation have been carried out. The 
results obtained in the present and in some earlier works are discussed 
taking into account existence of a wide component of lines, nondegenerative 
four photon mixing (NFPM) and time invariance violation in  the 
photon-molecule interactions. It is proposed to use the NFPM effect in 
investigations of the shape of  the optical transition in polyatomic molecules 
and especially for evaluation of  the homogeneous width of  the wide 
component of lines. The need and problems of experimental study of existing 
effects of  the T-invariance violation in optics are discussed.  

\vspace{5 pt}
\end{abstract}

\vspace{12 pt}

{\bf Introduction}

	Studies of  the IR-IR double resonance in $SF_6$ were widely carried 
out in 70's with a purpose to clarify  the mechanism of  the IR multiple 
photon excitation (IR MPE) of the polyatomic molecules. Large experimental 
material was obtained in this area in the subsequent years [1,2].  Also, a 
substantial progress was achieved  in  understanding the underlying physical 
processes.  All this demands to reexamine  the old experiments, as they 
contain  interesting and important information.

	A very strong bleaching in $SF_6$ was observed in the work [3]. This 
bleaching  was destroyed  by molecular collisions at  characteristic rate of  
$p\tau =48 $ns mbar. An effect of the direct multiphoton absorption of the 
pump laser radiation by molecules was proposed as a  candidate for explanation 
of this bleaching. On the contrary, in works [4,5] a hypothesis was proposed 
of existence of some wide component in the  absorption line of the polyatomic 
molecules. If the natural width of this wide component greater, than 
anharmonicity of vibrations, the IR MPE can occur in a natural way. Later, 
this idea was proved experimentally, when the far wings in the absorption 
band of polyatomic molecules were found [6], which were independent on the 
gas pressure. The intensities of these lorentzian-like band wings well 
coordinate with the saturation data of molecular absorption at low pressure, 
with the data of depletion of rotational levels in the $SF_6$ molecules [7,8] 
and with the data on the $SF_6$ absorption of laser radiation in the molecular 
beam conditions [9].  Integrated intensity of this wide component can be 
rather easily estimated from the saturated spectrum data. But the natural 
width can be measured correctly only in the molecular beam conditions, where 
rotational temperature is extremely low. Therefore, development of methods 
for measuring this width  in the room temperature cell conditions is of great 
interest. 

	Experiments on the IR-IR double resonance provide such an opportunity.
 The bleaching, observed in work [3], is very similar to the coherent effect 
of  NFPM.  In the latter case, redistribution of energy between a powerful 
pump beam and a weak probe beam takes place. It is well known, that such a 
process proceeds in $SF_6$ very  effectively (up to depletion of the pump 
beam) as a superradiation in the case of degenerative four photon mixing 
(DFPM) in a somewhat specific conditions (a short cell and high $SF_6$ 
pressure) [10]. As in the case of the coherent Raman gain spectroscopy 
[11,12], spectral dependence of the NFPM efficiency contains information about 
the shape of the optical transition. The peculiarity of this case is that the 
NFPM effect occurs within a very wide optical transition. Thus, the spectral 
width of the bleaching effect characterizes the homogeneous width of the 
optical transition.

	The purpose of the current work was to repeat the experiments of work 
[3], to perform some more detailed study of the spectral characteristics of  
the bleaching effect, to discuss  the role of  the NFPM and to propose to use 
it for determination of the widths of line wings of polyatomic molecules in 
the condition of cell experiments at room temperature. However, in the present 
experiments any nondegenerative bleaching was absent.

{\bf Experiment}

	The experimental set-up is shown in Fig.1. The main difference from 
work [3] was that a separate mirror (3) with R=-2m was used for focusing the 
laser radiation in the copper waveguide absorption cell (\o 10mm and length 
110cm). The probe laser radiation passed through the hole \o 8mm in the mirror 
(4) without focusing and was registered by an integrated calorimeter 2. Such 
an experimental set-up allows to work without a monochromator and to measure 
absorption of the probe laser radiation in the case of coinciding frequencies 
of the pump and the probe lasers. The difference in divergences of the 
scattered pump radiation and that of the probe laser is so great, that total 
energy of pump radiation, registered by the calorimeter 2 at the distance 
$\sim 1m$  from the mirror (4) is $<10 \mu j$ and does not prevent to measure 
the energy of the probe pulse.

	The resonator of the probe laser was formed by a rather flat mirror 
(R=-50m) and a diffraction grating. Its radiation was characterized by small 
divergence and good uniformity of the beam. Nitrogen free gas mixture was 
used. Multimode laser pulse had a full width about 100ns at half maximum 
without any "tail" and consisted of chaotic spikes.

	The pump laser had an unstable resonator and two very old $BaF_2$ 
coated NaCl windows. In the far field region, where it was focused, radiation 
was very non-uniform and had a typical speckle-structure. With small contents 
of nitrogen in the gas mixture, the multimode pulse consisted of a spike about 
100ns FWHM and a "tail" about $3\mu s$ full length, that contained about 20\% 
of the total pulse energy.  Directions of polarization of both laser beams 
coincided. Temporal characteristics were measured by means of a photon-drag 
detector and an oscilloscope (not shown in the figure).

The measurements were carried out at pressures 1,5 Torr $SF_6$ for the P10---
P14 and R10---R20 lines, 1,0 Torr for line P16  and 0,4 Torr $SF_6$ for lines 
P18---P30.

{\bf Experimental results}

The results of the IR-IR double resonance experiments are presented in 
Fig.2. The pump laser was tuned in the P16 line. Similar results were obtained 
when the pump laser was tuned in the P20 line. Incident pump laser fluence in 
the waveguide was $250 mj/cm^2$. Energy fluence of the probe laser radiation 
was $7 mj/cm^2$. The latter was decreased down to $1mj/cm^2$ in some 
experiments without any changes in the results.  In the whole range of 
testing (including the R10---R20 area) only increase in absorption of the 
probe laser radiation was observed. 

Some bleaching effect was observed only when both lasers were working on 
the same line. Apparent mutual influence of the lasers was observed in this 
case. Were the pulses of both lasers synchronized, energy of the probe laser 
pulse decreased by $\sim 10\% $, and its beam pattern became non-uniform. The 
distance between two lasers was about 10m in this case, so that only 
scattered pump laser radiation could get into the resonator of the probe laser.

	Also, a set of experiments was carried out in which the same laser 
radiation was used both for pumping and for probing. Again, the bleaching 
effect was found in this case. An open square in Fig.2 corresponds to the 
case, when the probe laser was used,  and the full square illustrates the 
case of using  only the pump laser radiation.

{\bf Discussion}

IR-IR double resonance was studied in numerous works [3,13-17]. These 
experiments frequently gave contradictory results, and it was difficult to 
give satisfactory explanations to these contradictions. At present, however, 
the situation has substantially cleared up. 

A wide component with Lorentzian FWHM $\sim 150GHz $ exists in the spectrum of 
narrow absorption lines of $SF_6$ molecule [6]. Its integral intensity for the 
$1\leftarrow 0$ transition of the $\nu_3 $ band contains only $\sim 0,2\% $ of 
the integral intensity of this optical transition. However, absorption of the 
powerful pump laser radiation takes place just through this component. The 
weaker probe radiation can be absorbed mainly through the narrow or through 
the wide component of the line. Thus, the results of probing can be 
different. All experiments on the IR-IR double resonance in $ SF_6 $ can be 
conditionally broken into three classes:  (1) when the probing occurs through 
the narrow component of lines, (2) when the probing occurs through the wide 
component of lines and (3) an intermediate case.

In the work [13] very weak probe radiation of the CW $CO_2$ laser with 
intensity $ \sim 1mW/cm^2 $ was used. Both induced transmissions in the 
P8---P22 region of the probe laser frequencies and induced absorption in the 
red side were observed. Saturation of the narrow component of lines was 
absent in this case. The main cause  of the observed effects was an 
anharmonic shift of the optical transitions of the excited molecules.

It is surprising, but experiments with megawatt intensity and picosecond 
timescale probe pulse laser radiation belong to the same class [14]. In this 
case  laser radiation has large spectral width ($\sim 1cm^{-1} $). Therefore, 
any saturation of the narrow component of lines is again absent, and the 
results of these experiments coincide practically with those of work [13]. 

CW $CO_2$ probe laser radiation  with intensity about $0,1-0,5 W/cm^2 $ was 
used in works [15-17]. These experiments belong to the intermediate class. 
Here we can see appearance of appreciable saturation of the narrow component 
of lines. Also, absorption through the wide component of lines begins to play 
some role. This role depends on exact tuning of the probe laser radiation 
frequency: is it set-up on a strong rotational line or between lines. Thus, 
in the area P8-P22 both induced transmission and induced absorption can be 
observed.

A specific effect of  short-time induced absorption in the course of the 
pump pulse was observed in work [16]. One should keep in mind, that planes of 
polarization of the probe and the pump laser radiation were orthogonal in this 
experiment. A coherent polarization effect could take place in this case,  
that is due to the population transfer between degenerated levels. Phenomena 
of this kind are studied in the atomic polarization spectroscopy [18],  and 
they are out of scope of the present paper.

Experiments of the present paper belong to the next class, when probing  
occurs mainly through the wide component of lines. Average probe laser 
intensity was $10-70 kW/cm^2$. Saturation of the narrow component of lines 
was much stronger in this case. The role of the wide component in absorption 
of the probing radiation became essential. 

In this case, besides the anharmonic shift,  an effect of sharp growth of 
integral intensity of the wide component in the region of low  lying 
vibrational levels should be taken into account. This effect manifests 
itself through a strong temperature dependence of the IR MPE process [9,19]. 
In some range of the experimental conditions this effect becomes decisive, 
and then induced absorption of the probe laser radiation can be observed in 
all spectral intervals.

To the same class of experiments the works [9,20] belong. In the work [9] 
the experiments were carried out in the molecular beam conditions. Along with 
rather high intensity of the probe laser radiation ($\sim 10kW/cm^2$), these 
experimental conditions were characterized by ultralow ($\sim 5^0 K$) 
rotational temperature [21].  A number of rotational lines were very low, and 
the probe laser radiation interacted practically with the wide component of 
lines only. The obtained results were similar to those of the present work. 
One should  keep in mind, however, that experiments of work [9] were carried 
out with rather hot gas ($T=500^0 K $). 

A special case is work [3], where, as we believe, a coherent NFPM effect 
was observed. The first argument for such a conclusion is unusually strong 
effect of collisions. Even elastic collisions can destroy coherency. The 
second argument is amplification in the region R12-R20,  that clearly 
indicates the combination $\nu_2 + \nu_6$ band can contribute to this process. 
In this case the phase-matched radiation must exist in the region near 
$915cm^{-1} $, and it is better to search for it in the direct beams without  
waveguide. And the third argument is that in our experiments, which 
were carried out in rather similar conditions, the characteristic bleaching 
was absent. This fact can be due to rather poor coherency of the pump 
beam in our case. It is known, that efficiency of the DFPM effect is 
extremely sensitive to the divergence and uniformity of the laser beam 
[10,22,23]. In the works [15-17] the NFPM effect might be absent as a 
result of low intensity of probe radiation, and in the experiments with 
molecular beam [9] its absence could be a consequence of low density and 
small size of  the interaction zone.

Is the bleaching effect, observed in our work, when both lasers radiate on 
the same line, a result of four photon mixing?  Absence of such an effect 
in the molecular beam conditions ( [9], Fig.3, without exact adjusting) is 
an indirect argument for the positive answer. In this case, the NFPM effect 
can be a consequence of a very small difference in the frequencies of  the 
pump and probe radiation.

Thus, detailed experimental studies of occurrence of the NFPM effect in 
$SF_6$ are of undoubted interest. They can give information about the shape of 
optical transitions, and not only about the forward transitions, but maybe 
about the backward ones.

Why one needs the  NFPM effect to present  within the wide and homogeneous 
optical transition?  Here we are concerning  a very important and interesting 
question about the time invariance violation in the photon-atom and photon-
molecule interactions. Basing on the considerations of symmetry, the theorists 
assume the existence of  the T-invariance violation effects in optics for a 
long time. It is not known, however, in which part of optics and in which 
form could these effects exist. For many years the scientists try to find 
them in the form of the so-called electrical dipole moment (EDM) [24]. The 
EDM is not found till now. 

General situation, however, is that the whole nonlinear optics based on or 
connected with T-invariance violation. {\bf The principle of T-invariance 
violation is obviously a quantum analog of the nonlinear susceptibility 
principle.} Most of the nonlinear effects in optics can be explained from 
the quantum point of view on the basis of the T-invariance violation 
principle. The effect of T-invariance violation usually manifests itself 
only in an indirect way, for example, as a population transfer effect at 
sweeping the resonance conditions [25]. Usually, experimenters cannot 
control the natural width and the cross-section of optical transitions in 
such a way to be able to compare these parameters for the forward and 
backward processes.

Maybe a unique exception for this day is the wide component of lines. It 
is a unique physical object that is characterized by an unusual combination of 
properties: a huge homogeneous width of transition is combined with long 
lifetime of the excited state. This allows to divorce in time and in space the 
process of laser excitation and that of monitoring the exited states of 
molecules [9]. These experiments yield fantastic results: the width of the 
backward optical transition proves to be more than five orders of magnitude 
less than that of  the forward one, the cross-section of the backward 
transition being more, than four orders of magnitudes greater, than that of 
the forward one [26]. These experiments provide a direct and full 
experimental proof of existence of  the strong T-invariance violation in  
the photon-molecule interactions. For more detailed experimental study of  
the T-invariance violation effect we need to reanimate and  
straightforwardly  continue experiments of work [9] (fig.5, 6).

In principle,  experiments of this type can be carried out in the cell 
conditions also. It is possible to study the dependence of absorption of the 
probe radiation on the angle between the probe  and the pump beams. We expect 
to see a transition from  direct stimulated deexcitation of molecules to the 
four photon mixing effect, which can reach a superradiation regime in optimum 
conditions [10,22]. We hope to prolong such experiments in future  using  the 
straight crossed beams without waveguide.  It would be useful, if  the same 
studies were conducted somewhere else. 

The author is grateful to Dr. V.S.Mezhevov for equipping us with  a pump 
laser and to Dr. A.P.Dyad'kin for discussions.

\end{document}